\documentclass[12pt,a4paper]{article}

\usepackage{ifthen} 
\newboolean{pdflatex}
\setboolean{pdflatex}{true} 

\newboolean{articletitles}
\setboolean{articletitles}{true} 

\newboolean{uprightparticles}
\setboolean{uprightparticles}{false} 

\usepackage{lineno}  
\usepackage{xspace} 

\usepackage{graphicx}  
\usepackage{color}
\usepackage{colortbl}
\graphicspath{{./figs/}} 

\usepackage{amsmath} 
\usepackage{amssymb}
\usepackage{amsfonts}
\usepackage{upgreek} 

\newcommand*\patchAmsMathEnvironmentForLineno[1]{%
\expandafter\let\csname old#1\expandafter\endcsname\csname #1\endcsname
\expandafter\let\csname oldend#1\expandafter\endcsname\csname
end#1\endcsname
 \renewenvironment{#1}%
   {\linenomath\csname old#1\endcsname}%
   {\csname oldend#1\endcsname\endlinenomath}%
}
\newcommand*\patchBothAmsMathEnvironmentsForLineno[1]{%
  \patchAmsMathEnvironmentForLineno{#1}%
  \patchAmsMathEnvironmentForLineno{#1*}%
}
\AtBeginDocument{%
\patchBothAmsMathEnvironmentsForLineno{equation}%
\patchBothAmsMathEnvironmentsForLineno{align}%
\patchBothAmsMathEnvironmentsForLineno{flalign}%
\patchBothAmsMathEnvironmentsForLineno{alignat}%
\patchBothAmsMathEnvironmentsForLineno{gather}%
\patchBothAmsMathEnvironmentsForLineno{multline}%
}

\usepackage{hyperref}    
\usepackage[all]{hypcap} 

\def\lhcb {\mbox{LHCb}\xspace}
\def\ux85 {\mbox{UX85}\xspace}

\def\cdf    {\mbox{CDF}\xspace}
\def\dzero  {\mbox{D0}\xspace}

\def\tevatron {Tevatron\xspace}

\ifthenelse{\boolean{uprightparticles}}%
{
 
 \def\Pgamma      {\ensuremath{\upgamma}\xspace}

 \def\Pmu         {\ensuremath{\upmu}\xspace}

 \def\Ppi         {\ensuremath{\uppi}\xspace}

 \def\Ppsi        {\ensuremath{\uppsi}\xspace}

 \def\PDelta      {\ensuremath{\Delta}\xspace}                 
 \def\PXi      {\ensuremath{\Xi}\xspace}                 
 \def\PLambda      {\ensuremath{\Lambda}\xspace}                 
 \def\PSigma      {\ensuremath{\Sigma}\xspace}                 
 \def\POmega      {\ensuremath{\Omega}\xspace}                 
 \def\PUpsilon      {\ensuremath{\Upsilon}\xspace}                 
 
 \def\PB      {\ensuremath{\mathrm{B}}\xspace}                 
                  
 \def\PD      {\ensuremath{\mathrm{D}}\xspace}

 \def\PJ      {\ensuremath{\mathrm{J}}\xspace}                 
 \def\PK      {\ensuremath{\mathrm{K}}\xspace}

 \def\Pb      {\ensuremath{\mathrm{b}}\xspace}                 
 \def\Pc      {\ensuremath{\mathrm{c}}\xspace}                 
 \def\Pd      {\ensuremath{\mathrm{d}}\xspace}

 \def\Pi      {\ensuremath{\mathrm{i}}\xspace}

 \def\Pp      {\ensuremath{\mathrm{p}}\xspace}                 
 \def\Pq      {\ensuremath{\mathrm{q}}\xspace}                 
                  
 \def\Ps      {\ensuremath{\mathrm{s}}\xspace}                 
                  
 \def\Pu      {\ensuremath{\mathrm{u}}\xspace}

}
{
 
 \def\Pgamma      {\ensuremath{\gamma}\xspace}

 \def\Pmu         {\ensuremath{\mu}\xspace}

 \def\Ppi         {\ensuremath{\pi}\xspace}

 \def\Ppsi        {\ensuremath{\psi}\xspace}                 
                  
 \mathchardef\PDelta="7101
 \mathchardef\PXi="7104
 \mathchardef\PLambda="7103
 \mathchardef\PSigma="7106
 \mathchardef\POmega="710A
 \mathchardef\PUpsilon="7107
                  
 \def\PB      {\ensuremath{B}\xspace}                 
                  
 \def\PD      {\ensuremath{D}\xspace}

 \def\PJ      {\ensuremath{J}\xspace}                 
 \def\PK      {\ensuremath{K}\xspace}

 \def\Pb      {\ensuremath{b}\xspace}                 
 \def\Pc      {\ensuremath{c}\xspace}                 
 \def\Pd      {\ensuremath{d}\xspace}

 \def\Pi      {\ensuremath{i}\xspace}

 \def\Pp      {\ensuremath{p}\xspace}                 
 \def\Pq      {\ensuremath{q}\xspace}                 
                  
 \def\Ps      {\ensuremath{s}\xspace}                 
                  
 \def\Pu      {\ensuremath{u}\xspace}

}

\def\mup        {\ensuremath{\Pmu^+}\xspace}
\def\mun        {\ensuremath{\Pmu^-}\xspace} 

\def\g      {\ensuremath{\Pgamma}\xspace}

\def\quark     {\ensuremath{\Pq}\xspace}

\def\uquark    {\ensuremath{\Pu}\xspace}

\def\dquark    {\ensuremath{\Pd}\xspace}

\def\squark    {\ensuremath{\Ps}\xspace}

\def\cquark    {\ensuremath{\Pc}\xspace}

\def\bquark    {\ensuremath{\Pb}\xspace}

\def\pion  {\ensuremath{\Ppi}\xspace}

\def\pip   {\ensuremath{\pion^+}\xspace}
\def\pim   {\ensuremath{\pion^-}\xspace}
\def\pipi  {\ensuremath{\pion^+\pion^-}\xspace}
\def\pipm  {\ensuremath{\pion^\pm}\xspace}

\def\kaon  {\ensuremath{\PK}\xspace}
  \def\Kbar  {\kern 0.2em\overline{\kern -0.2em \PK}{}\xspace}

\def\Kz    {\ensuremath{\kaon^0}\xspace}
\def\Kzb   {\ensuremath{\Kbar^0}\xspace}
\def\KzKzb {\ensuremath{\Kz \kern -0.16em \Kzb}\xspace}
\def\Kp    {\ensuremath{\kaon^+}\xspace}
\def\Km    {\ensuremath{\kaon^-}\xspace}

\def\KpKm  {\ensuremath{\Kp \kern -0.16em \Km}\xspace}
\def\KS    {\ensuremath{\kaon^0_{\rm\scriptscriptstyle S}}\xspace}

  \def\Dbar    {\kern 0.2em\overline{\kern -0.2em \PD}{}\xspace}
\def\D       {\ensuremath{\PD}\xspace}

\def\Dz      {\ensuremath{\D^0}\xspace}
\def\Dzb     {\ensuremath{\Dbar^0}\xspace}
\def\DzDzb   {\ensuremath{\Dz {\kern -0.16em \Dzb}}\xspace}
\def\Dp      {\ensuremath{\D^+}\xspace}
\def\Dm      {\ensuremath{\D^-}\xspace}

\def\DpDm    {\ensuremath{\Dp {\kern -0.16em \Dm}}\xspace}

\def\Dstarp  {\ensuremath{\D^{*+}}\xspace}

  \def\Bbar    {\kern 0.18em\overline{\kern -0.18em \PB}{}\xspace}

\def\jpsi     {\ensuremath{{\PJ\mskip -3mu/\mskip -2mu\Ppsi\mskip 2mu}}\xspace}
\def\psitwos  {\ensuremath{\Ppsi{(2S)}}\xspace}

  \def\Y#1S{\ensuremath{\PUpsilon{(#1S)}}\xspace}

\def\proton      {\ensuremath{\Pp}\xspace}

\def\Xires {\ensuremath{\PXi}\xspace}

\def\L {\ensuremath{\PLambda}\xspace}
\def\Lbar {\ensuremath{\kern 0.1em\overline{\kern -0.1em\PLambda}}\xspace}

\def\Sigmares {\ensuremath{\PSigma}\xspace}

\def\Omegares {\ensuremath{\POmega}\xspace}

\def\Lb      {\ensuremath{\L^0_\bquark}\xspace}

\def\Lc      {\ensuremath{\L^+_\cquark}\xspace}

\newcommand{\decay}[2]{\ensuremath{#1\!\to #2}\xspace}         

\def\to                 {\ensuremath{\rightarrow}\xspace}

\def\AT#1     {\ensuremath{A_{\mathrm{T}}^{#1}}\xspace}           

\def\C#1      {\ensuremath{\mathcal{C}_{#1}}\xspace}                       
\def\Cp#1     {\ensuremath{\mathcal{C}_{#1}^{'}}\xspace}                    
\def\Ceff#1   {\ensuremath{\mathcal{C}_{#1}^{\mathrm{(eff)}}}\xspace}        
\def\Cpeff#1  {\ensuremath{\mathcal{C}_{#1}^{'\mathrm{(eff)}}}\xspace}       
\def\Ope#1    {\ensuremath{\mathcal{O}_{#1}}\xspace}                       
\def\Opep#1   {\ensuremath{\mathcal{O}_{#1}^{'}}\xspace}                    

\newcommand{\tev}{\ensuremath{\mathrm{\,Te\kern -0.1em V}}\xspace}
\newcommand{\gev}{\ensuremath{\mathrm{\,Ge\kern -0.1em V}}\xspace}
\newcommand{\mev}{\ensuremath{\mathrm{\,Me\kern -0.1em V}}\xspace}
\newcommand{\kev}{\ensuremath{\mathrm{\,ke\kern -0.1em V}}\xspace}
\newcommand{\ev}{\ensuremath{\mathrm{\,e\kern -0.1em V}}\xspace}
\newcommand{\gevc}{\ensuremath{{\mathrm{\,Ge\kern -0.1em V\!/}c}}\xspace}
\newcommand{\mevc}{\ensuremath{{\mathrm{\,Me\kern -0.1em V\!/}c}}\xspace}
\newcommand{\gevcc}{\ensuremath{{\mathrm{\,Ge\kern -0.1em V\!/}c^2}}\xspace}
\newcommand{\gevgevcccc}{\ensuremath{{\mathrm{\,Ge\kern -0.1em V^2\!/}c^4}}\xspace}
\newcommand{\mevcc}{\ensuremath{{\mathrm{\,Me\kern -0.1em V\!/}c^2}}\xspace}

\def\mum  {\ensuremath{\,\upmu\rm m}\xspace}

\def\invfb   {\ensuremath{\mbox{\,fb}^{-1}}\xspace}

\newcommand{\chisq}{\ensuremath{\chi^2}\xspace}

\def\gsim{{~\raise.15em\hbox{$>$}\kern-.85em
          \lower.35em\hbox{$\sim$}~}\xspace}
\def\lsim{{~\raise.15em\hbox{$<$}\kern-.85em
          \lower.35em\hbox{$\sim$}~}\xspace}

\def\sWeights{\mbox{\em sWeights}\xspace}

\def\pt         {\mbox{$p_{\rm T}$}\xspace}

\def\dllkpi     {\ensuremath{\mathrm{DLL}_{\kaon\pion}}\xspace}
\def\dllppi     {\ensuremath{\mathrm{DLL}_{\proton\pion}}\xspace}

\def\evtgen     {\mbox{\textsc{EvtGen}}\xspace}
\def\pythia     {\mbox{\textsc{Pythia}}\xspace}

\def\geant      {\mbox{\textsc{Geant4}}\xspace}

\def\photos     {\mbox{\textsc{Photos}}\xspace}

\def\tell1  {TELL1\xspace}
\def\ukl1   {UKL1\xspace}

\newcommand{\lbst}{\ensuremath{\L_{\bquark}^{*0}}\xspace}
\newcommand{\lbl}{\ensuremath{\L_{\bquark}^{*0}(5912)}\xspace}
\newcommand{\lbh}{\ensuremath{\L_{\bquark}^{*0}(5920)}\xspace}
\newcommand{\lblcpi}{\ensuremath{\decay{\Lb}{\Lc\pim}}\xspace}
\newcommand{\lblck}{\ensuremath{\decay{\Lb}{\Lc\Km}}\xspace}
\newcommand{\lcpi}{\ensuremath{\Lc\pim}\xspace}
\newcommand{\lck}{\ensuremath{\Lc\Km}\xspace}
\newcommand{\lcpkpi}{\ensuremath{\decay{\Lc}{\proton\Km\pip}}\xspace}
\newcommand{\lbpp}{\ensuremath{\Lb\pipi}\xspace}
\newcommand{\lbppss}{\ensuremath{\Lb\pipm\pipm}\xspace}
\newcommand{\lbg}{\ensuremath{\Lb\g}\xspace}

\usepackage{cite} 
\usepackage{mciteplus}

\begin{document}

\renewcommand{\thefootnote}{\fnsymbol{footnote}}
\setcounter{footnote}{1}


\begin{titlepage}

\pagenumbering{roman}

\vspace*{-1.5cm}
\centerline{\large EUROPEAN ORGANIZATION FOR NUCLEAR RESEARCH (CERN)}
\vspace*{1.5cm}
\hspace*{-0.5cm}
\begin{tabular*}{\linewidth}{lc@{\extracolsep{\fill}}r}
\ifthenelse{\boolean{pdflatex}}
{\vspace*{-2.7cm}\mbox{\!\!\!\includegraphics[width=.14\textwidth]{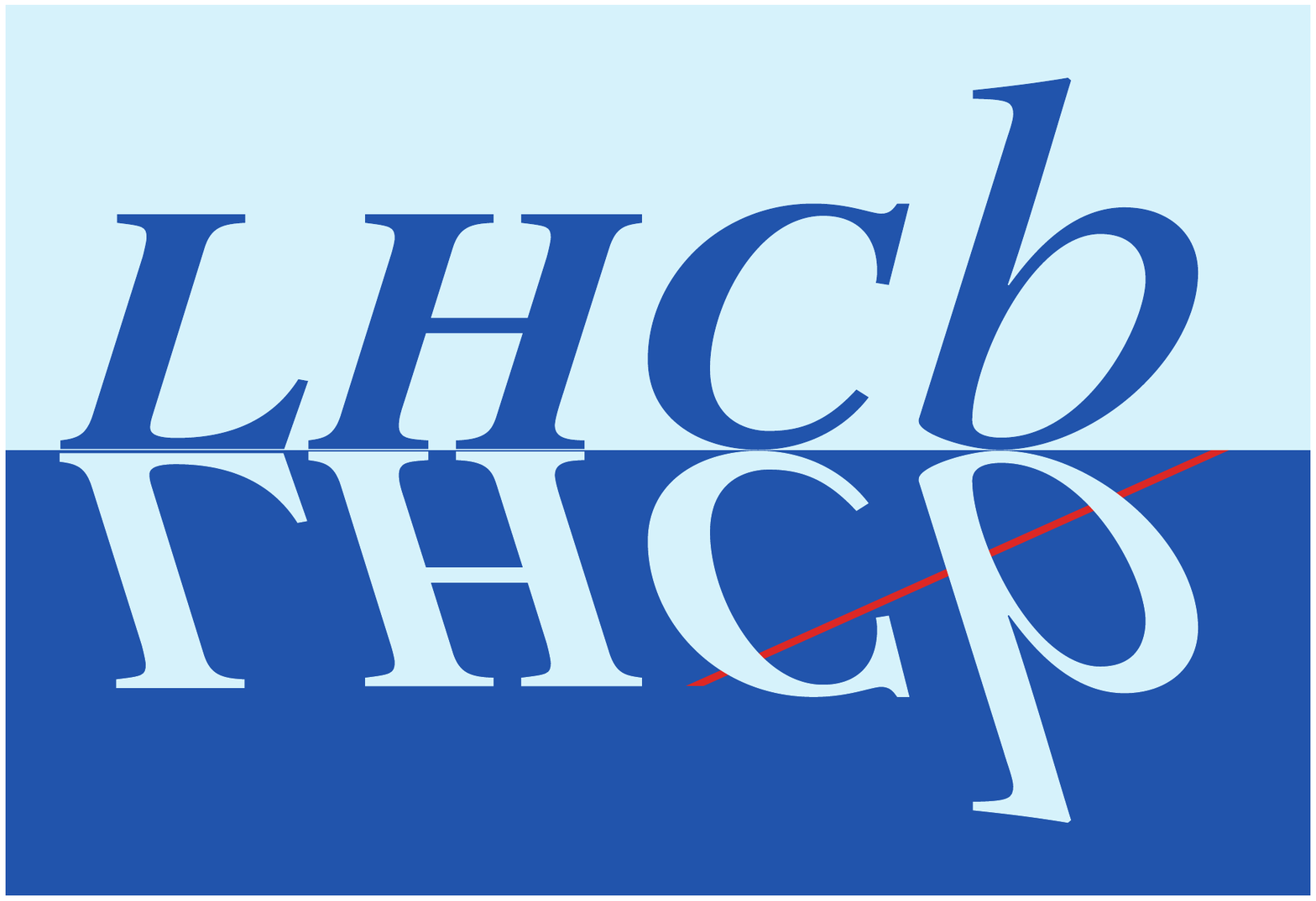}} & &}%
{\vspace*{-1.2cm}\mbox{\!\!\!\includegraphics[width=.12\textwidth]{figs/lhcb-logo.eps}} & &}%
\\
 & & CERN-PH-EP-2012-128 \\  
 & & LHCb-PAPER-2012-012 \\  
 & & 28 October 2012 \\ 
 & & \\
\end{tabular*}

\vspace*{4.0cm}

{\bf\boldmath\huge
\begin{center}
Observation of excited \boldmath{\Lb} baryons
\end{center}
}

\vspace*{2.0cm}

\begin{center}
The LHCb collaboration
\footnote{Authors are listed on the following pages.}
\end{center}

\vspace{\fill}

\begin{abstract}
  \noindent
  Using $\proton\proton$ collision data corresponding to 1.0~\invfb integrated luminosity
  collected by the \lhcb detector, two narrow states are observed in the \lbpp spectrum with masses 
  $5911.97\pm 0.12(\mbox{stat})\pm 0.02(\mbox{syst})\pm 0.66(\Lb\mbox{ mass})$ \mevcc and 
  $5919.77\pm 0.08(\mbox{stat})\pm 0.02(\mbox{syst})\pm 0.66(\Lb\mbox{ mass})$ \mevcc. 
  The significances of the observations are $5.2$ and $10.2$ standard deviations, 
  respectively. These states are interpreted as the orbitally excited \Lb baryons, $\lbl$ and $\lbh$. 
\end{abstract}

\vspace*{2.0cm}

\begin{center}
To be submitted to Phys. Rev. Lett.
\end{center}

\vspace{\fill}

\end{titlepage}


\newpage
\setcounter{page}{2}
\mbox{~}
\newpage

\centerline{\large\bf The LHCb collaboration}
\begin{flushleft}
\small
R.~Aaij$^{38}$, 
C.~Abellan~Beteta$^{33,n}$, 
A.~Adametz$^{11}$, 
B.~Adeva$^{34}$, 
M.~Adinolfi$^{43}$, 
C.~Adrover$^{6}$, 
A.~Affolder$^{49}$, 
Z.~Ajaltouni$^{5}$, 
J.~Albrecht$^{35}$, 
F.~Alessio$^{35}$
% ...

{\footnotesize \it
% ...
$ ^{5}$Clermont Universit\'{e}, Universit\'{e} Blaise Pascal, CNRS/IN2P3, LPC, Clermont-Ferrand, France\\
$ ^{6}$CPPM, Aix-Marseille Universit\'{e}, CNRS/IN2P3, Marseille, France\\
% ...
$ ^{11}$Physikalisches Institut, Ruprecht-Karls-Universit\"{a}t Heidelberg, Heidelberg, Germany\\
% ...
$ ^{33}$Universitat de Barcelona, Barcelona, Spain\\
$ ^{34}$Universidad de Santiago de Compostela, Santiago de Compostela, Spain\\
$ ^{35}$European Organization for Nuclear Research (CERN), Geneva, Switzerland\\
% ...
$ ^{38}$Nikhef National Institute for Subatomic Physics, Amsterdam, The Netherlands\\
% ...
$ ^{43}$H.H. Wills Physics Laboratory, University of Bristol, Bristol, United Kingdom\\
% ...
$ ^{49}$Oliver Lodge Laboratory, University of Liverpool, Liverpool, United Kingdom\\
% ...
$ ^{n}$LIFAELS, La Salle, Universitat Ramon Llull, Barcelona, Spain\\
% ...
}
\end{flushleft}

\cleardoublepage


\renewcommand{\thefootnote}{\arabic{footnote}}
\setcounter{footnote}{0}



\pagestyle{plain} 
\setcounter{page}{1}
\pagenumbering{arabic}

\linenumbers

%

\linenumbers

The system of baryons containing a \bquark quark (beauty baryons) remains largely unexplored, despite recent 
progress made at the experiments at the \tevatron. In addition to the ground state, \Lb, 
the $\Xires_{\bquark}^-$ baryon with the quark content $\bquark\squark\dquark$ has been observed by the 
\dzero~\cite{Abazov:2007ub} and \cdf~\cite{Aaltonen:2007un} collaborations, followed by the observation 
of the doubly-strange $\Omegares_{\bquark}^-$ baryon ($\bquark\squark\squark$) \cite{Abazov:2008qm, Aaltonen:2009ny}. 
The last ground state of beauty-strange content, $\Xires_{\bquark}^0$ ($\bquark\squark\uquark$), has been observed
by \cdf~\cite{Aaltonen:2011wd}. Recently, the CMS collaboration has found the corresponding excited state, 
most likely $\Xires_{\bquark}^{*0}$ with $J^P=3/2^+$~\cite{Chatrchyan:2012ni}. 
Beauty baryons with two light quarks ($\bquark\quark\quark$, where 
$\quark=\uquark,\dquark$), other than the \Lb, have been studied so far by \cdf
only. Of the triplets $\Sigmares_{\bquark}^{\pm,0}$ with spin $J=1/2$ and 
$\Sigmares_{\bquark}^{*\pm,0}$ with $J=3/2$ predicted by theory, only the charged states 
$\Sigmares_{\bquark}^{(*)\pm}$ have so far been observed via their decay 
to $\Lb\pipm$ final states~\cite{Aaltonen:2007rw, CDF:2011ac}. None of the quantum numbers of beauty
baryons have been measured. 

The quark model predicts the existence of two orbitally excited \Lb states, \lbst, with the quantum 
numbers $J^P=1/2^-$ and $3/2^-$, respectively, that should decay to $\lbpp$ or $\lbg$.
These states have not previously been established experimentally. The properties of excited \Lb baryons 
are discussed in Refs.~\cite{AzizaBaccouche:2001jc, AzizaBaccouche:2001pu, Ebert:2007nw, Garcilazo:2007eh, Capstick:1986bm, Karliner:2008sv, Roberts:2007ni}. Most predictions give masses above the $\lbpp$ threshold, but
below the $\Sigmares_{\bquark}\pion$ threshold. Observation of \lbst states and measurement of their 
quantum numbers would provide a further confirmation of the validity of the quark model, 
and the precise measurement of their masses would test the applicability of 
various theoretical models used to describe the interaction of heavy quarks. 

This Letter reports the first observation of the $\lbst$ states decaying into $\lbpp$, 
and the measurement of their masses and upper limits on their natural widths. 
The data set of 1.0~\invfb collected in $\proton\proton$ collisions at the LHC collider
at the center-of-mass energy $\sqrt{s}=7$~\tev in 2011 is used for the analysis. 

The \lhcb detector~\cite{Alves:2008zz} is a single-arm forward
spectrometer covering the pseudorapidity range $2<\eta <5$, designed
for the study of particles containing \bquark or \cquark quarks. The
detector includes a high precision tracking system consisting of a
silicon-strip vertex detector surrounding the $pp$ interaction region,
a large-area silicon-strip detector located upstream of a dipole
magnet with a bending power of about $4{\rm\,Tm}$, and three stations
of silicon-strip detectors and straw drift tubes placed
downstream. The combined tracking system has a momentum resolution
$\Delta p/p$ that varies from 0.4\% at 5\gevc to 0.6\% at 100\gevc,
and an impact parameter (IP) resolution of 20\mum for tracks with high
transverse momentum. Charged hadrons are identified using two
ring-imaging Cherenkov (RICH) detectors. Photon, electron and hadron
candidates are identified by a calorimeter system consisting of
scintillating-pad and preshower detectors, an electromagnetic
calorimeter and a hadronic calorimeter. Muons are identified by a muon
system composed of alternating layers of iron and multiwire
proportional chambers. 

The online event selection (trigger) consists of a hardware stage, based
on information from the calorimeter and muon systems, followed by a
software stage which applies full event reconstruction.
The software trigger used in this analysis requires a two-, three- or four-track
secondary vertex with a high sum of the momenta transverse to the beam axis, \pt, of
the tracks, and significant displacement from the primary interaction vertex (PV). 
In addition, the secondary vertex should have at least one track with $\pt > 1.7\gevc$, 
IP $\chi^2$ with respect to any PV greater than 16 (where the 
IP $\chi^2$ is defined as the difference of the PV fit $\chi^2$ with and without 
the track included),
and a track fit $\chisq/\rm{ndf} < 2$ where $\rm{ndf}$ is the number of 
degrees of freedom in the fit. A multivariate algorithm is used
for the identification of the secondary vertices~\cite{LHCb-PUB-2011-016}.

The \Lb candidates are reconstructed in the \lblcpi, \lcpkpi decay chain
(addition of charge-conjugate states is implied throughout this Letter). The 
selection of \Lb candidates is performed in two stages. First, a loose preselection 
of events containing beauty hadron candidates decaying to charm hadron candidates
is performed. It requires that the tracks forming the candidate, as well as the 
beauty and charm vertices, have good quality and are well separated from any 
PV, and the invariant masses of the beauty and charm candidates 
are consistent with the masses of the corresponding particles. 

The final selection requires that all the tracks forming the \Lb candidate have an 
IP $\chi^2$ with respect to any PV greater than 9, 
and the IP $\chi^2$ of the \Lb candidate to the best PV (PV having the minimum IP $\chi^2$ for the \Lb candidate) 
is less than 16. Particle identification (PID) information from the RICH 
detectors is used to identify kaons and protons in the final state in the form 
of differences of logarithms of likelihoods between the proton and pion 
(\dllppi) and kaon and pion (\dllkpi) hypotheses. 
No PID requirements are applied to the pions from \lblcpi decays to increase the \Lb yield: a
significant fraction of these pions have momenta above 100~\gevc where the PID performance is reduced. 
Finally, a kinematic fit 
is used which constrains the decay products of the $\Lb$ and $\Lc$ baryons to originate from 
common vertices, the $\Lb$ to originate from the PV and the invariant mass of the \Lc candidate
to be equal to the established \Lc mass~\cite{Nakamura:2010zzi}. 

A momentum scale correction is applied to all invariant mass spectra in this analysis 
to improve the mass measurement using the procedure similar to~\cite{lhcb_lb_mass}. 
The momentum scale has been calibrated using $\jpsi\to\mup\mun$
decays, and its accuracy has been quantified with other two-body resonance 
decays ($\Y1S\to\mup\mun$, $\KS\to\pipi$, $\phi\to\Kp\Km$).

Signal and background distributions are studied using simulation. 
Proton-proton collisions are generated using
\pythia~6.4~\cite{Sjostrand:2006za} with a specific \lhcb
configuration~\cite{LHCb-PROC-2010-056}.  Decays of hadronic particles
are described by \evtgen~\cite{Lange:2001uf} in which final state
radiation is generated using \photos~\cite{Golonka:2005pn}. The
interaction of the generated particles with the detector and its
response are implemented using the \geant
toolkit~\cite{Allison:2006ve, *Agostinelli:2002hh} as described in
Ref.~\cite{LHCb-PROC-2011-006}.

\begin{figure}
  \begin{center}
    \includegraphics[width=0.53\textwidth]{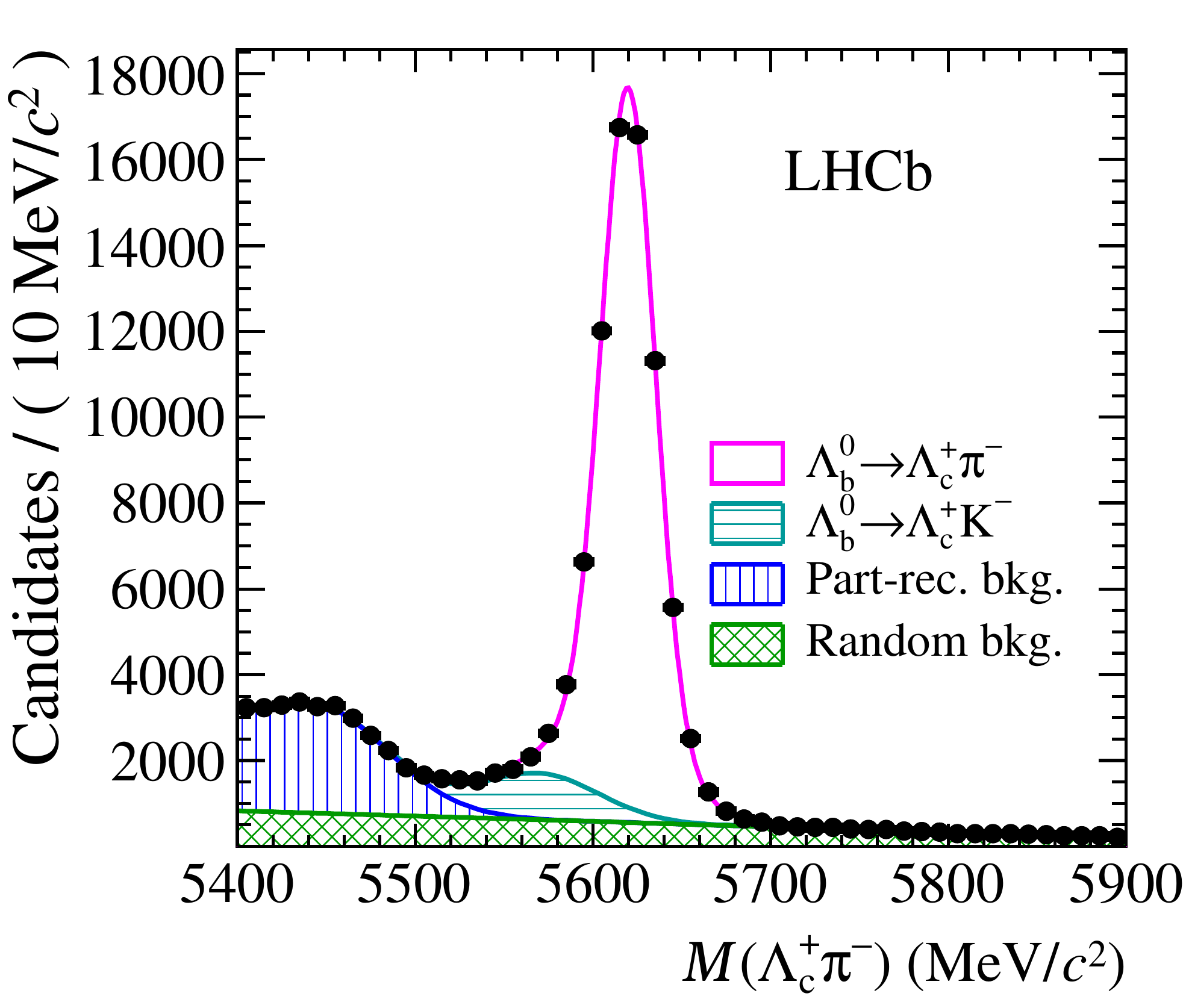}
  \end{center}
  \vspace*{-0.5cm}
  \caption{\small Invariant mass spectrum of \lcpi combinations. The points with error bars are the data, 
    and the fitted \lblcpi signal and three background components (\lblck, partially-reconstructed 
    and random background) are shown with different fill styles. }
  \label{fig:lb}
\end{figure}

The distribution of the \lcpi invariant mass after the kinematic fit is shown in 
Fig.~\ref{fig:lb}, where a requirement of good quality of the kinematic fit is applied. 
In addition to the \lblcpi signal contribution, the spectrum 
contains backgrounds from random combinations of tracks (random background), 
from partially-reconstructed decays where one or more particles are not reconstructed, 
and from \lblck decays with the kaon reconstructed under the pion mass hypothesis. 
A fit of the spectrum yields $70\,540\pm 330$ signal events, and the signal-to-background 
ratio in a $\pm 25$~\mevcc interval around the nominal \Lb mass is $S/B=11$. The fit to the
\lcpi spectrum is only used to estimate the \Lb yield and the $\Lb\to\lck$ 
contribution, and is not used in the subsequent analysis. 

The \Lb candidates obtained with the above selection are combined with two tracks under the pion mass 
hypothesis (referred to as slow pions from now on) to search for excited \Lb states. 
The tracks are required to have transverse momentum 
$\pt>150 \mevc$, and no PID requirements are applied. A kinematic fit is applied that, in addition 
to all constraints described above for \Lb candidates, constrains the two slow pion tracks to originate from the PV
and the invariant mass of the \Lb candidate to a fixed value of $5619.37$~\mevcc, which is 
a combination of the world average~\cite{Nakamura:2010zzi} and the LHCb measurement~\cite{LHCb-PAPER-2011-035}. 
The uncertainty on the combined $\Lb$ mass obtained in this way, $0.69$~\mevcc, is treated as a 
systematic effect. 
Combinations with a good quality of kinematic fit, $\chi^2/{\rm ndf}<3.3$, are retained. From the 
simulation study, this requirement is optimal for the observation of a narrow state near the kinematic 
threshold with signal-to-background ratio around one. 

The fit of the \lcpi mass spectrum (Fig.~\ref{fig:lb}) indicates the presence of the background from 
\lblck decays at a rate around 12\%, relative to the \lblcpi signal. 
Alternatively, its rate can be estimated from the ratio of $B^+\to \overline{D}{}^0K^+$ and 
$B^+\to \overline{D}{}^0\pi^+$ decays that equals to 8\%~\cite{Nakamura:2010zzi}. 
Due to the \Lb mass constraint in the kinematic fit, the \lbpp invariant mass distribution for this mode is 
biased by less than 0.1~\mevcc if reconstructed under the \lcpi mass hypothesis, and has a resolution only a factor of 
two worse than that with the \lcpi signal. After the kinematic fit quality requirement, the fraction of \lbpp with \lblck 
decays compared to those with the \lcpi is reduced to 8\%. This mode is thus not treated separately, and its 
effect is taken into account as a part of the systematic uncertainty due to the signal shape. 

\begin{figure}
  \begin{center}
    \includegraphics[width=0.67\textwidth]{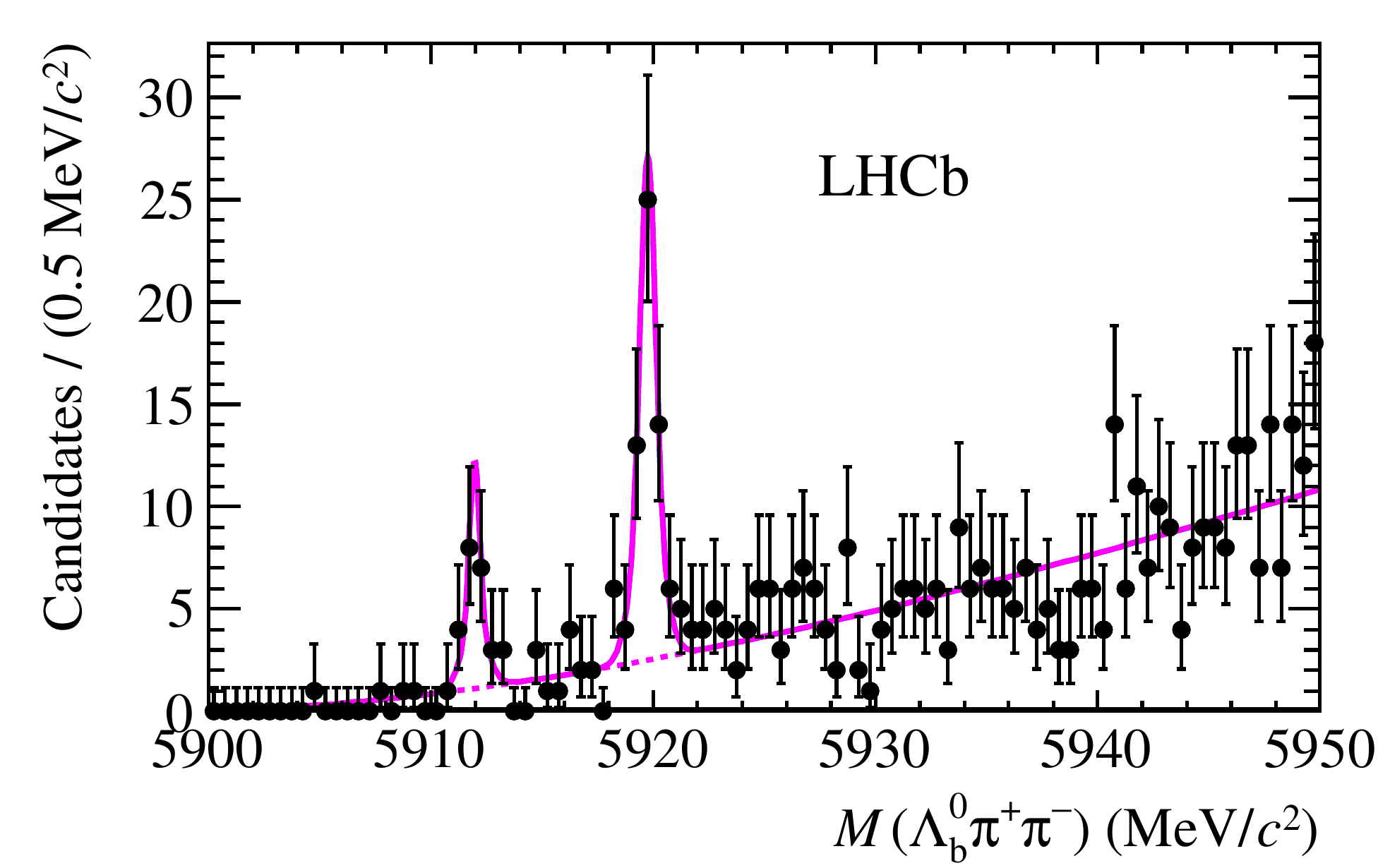}
    \put(-230,150){(a)}
    \\
    \includegraphics[width=0.67\textwidth]{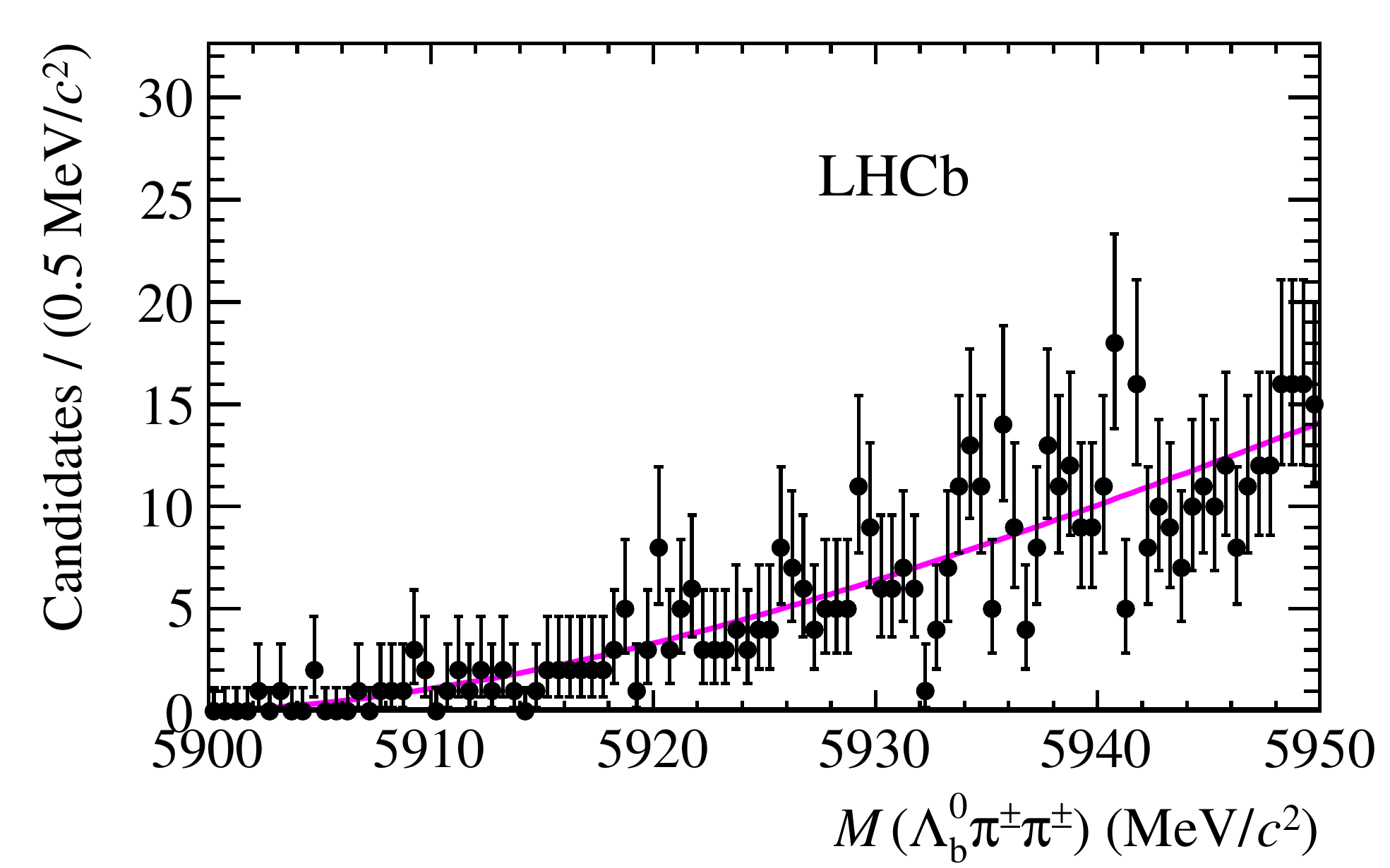}
    \put(-230,150){(b)}
  \end{center}
  \vspace*{-0.5cm}
  \caption{\small Invariant mass spectrum of (a) \lbpp and (b) \lbppss combinations. 
      The points with error bars are the data, 
      the solid line is the fit result, the dashed line is the background contribution. }
  \label{fig:lbstar}
\end{figure}

Combinations of \Lb candidates with both opposite-sign and same-sign slow pions are selected in data. 
The latter are used to constrain the background shape coming from random combinations of \Lb baryon and two tracks. 
The assumption that the shape of the background in \lbpp and \lbppss modes is the same is validated with simulation. 
The \lbpp and \lbppss invariant mass spectra are shown in Fig.~\ref{fig:lbstar}; two narrow structures with masses 
around 5912 and 5920~\mevcc are evident in the \lbpp spectrum. They are interpreted as the orbitally excited \Lb states, and are denoted hereafter as \lbl and \lbh. 

A combined unbinned fit of the \lbpp and \lbppss samples is performed 
to extract the masses and event yields of the two states. The background is described with a quadratic polynomial
function with common parameters for both samples except for an overall normalization. The probability density function 
(PDF) for each  of the \lbl and \lbh signals is a sum of two Gaussian PDFs with the same mean. 
The relative normalization of the two Gaussian PDFs are fixed to the values obtained from the simulation of 
states with masses 5912 and 5920~\mevcc and zero natural widths, while the mean value and overall normalization for each
signal are left free in the fit. The core resolution (width of the narrower Gaussian PDF) obtained from simulation is 0.19 
and 0.27~\mevcc for \lbl and \lbh, respectively. Study of several high-statistics samples ($\lblcpi$, $\psitwos\to\jpsi\pipi$, 
$\Dstarp\to\Dz\pip$) shows that the invariant mass resolution in data is typically worse by $20\%$ than in the simulation.
Thus the nominal data fit uses the widths of Gaussian PDFs from the simulation multiplied by 1.2. The data fit 
yields $17.6\pm 4.8$ events with mass $M_{\lbl}=5911.97\pm 0.12$~\mevcc and $52.5\pm 8.1$ events with 
mass $M_{\lbh}=5919.77\pm 0.08$~\mevcc. 

Limits on natural widths $\Gamma$ of the two states are obtained by performing an alternative fit where the signal PDFs 
are convolved with relativistic Breit-Wigner distributions. 
The dependence of Breit-Wigner width $\Gamma$ on the $\Lambda_b^0\pi^+\pi^-$ invariant mass $M$
is taken into account as $\Gamma_{\lbst}(M) = \Gamma_{\lbst}\times(q/q_0)^2\times(M_{\lbst}/M)$. 
Here $M_{\lbst}$ is the mass of the \lbst state, and $q_{(0)}$ is the kinematic energy for the decay 
of the state with mass $M_{(\lbst)}$: $q_{(0)}=M_{(\lbst)}-M_{\Lb}-2M_{\pi}$, where $M_{\Lb}$ and 
$M_{\pi}$ are the masses of $\Lb$ and $\pip$, respectively. Scans of Breit-Wigner widths 
$\Gamma_{\lbl}$ and $\Gamma_{\lbh}$ are performed with all the other parameters free to vary in the fit. 
The upper limits are obtained without applying the mass resolution scaling factor of 1.2 as in the nominal 
fit to account for the uncertainty of this quantity: this gives a more conservative value for the upper limit. 
The 90\% (95\%) confidence level (CL) upper limit on $\Gamma$, which corresponds to 1.28 (1.64) standard deviations, 
is obtained as the value of $\Gamma$ where the negative logarithm of the likelihood is
$1.28^2/2=0.82$ ($1.64^2/2=1.34$) greater than at its minimum. 
The 90\% (95\%) CL upper limit is $\Gamma_{\lbl}<0.66$~\mev 
($0.83$~\mev) for the \lbl state, and $\Gamma_{\lbh}<0.63$~\mev ($0.75$~\mev) for the \lbh state.

\begin{figure}
  \begin{center}
    \includegraphics[width=0.47\textwidth]{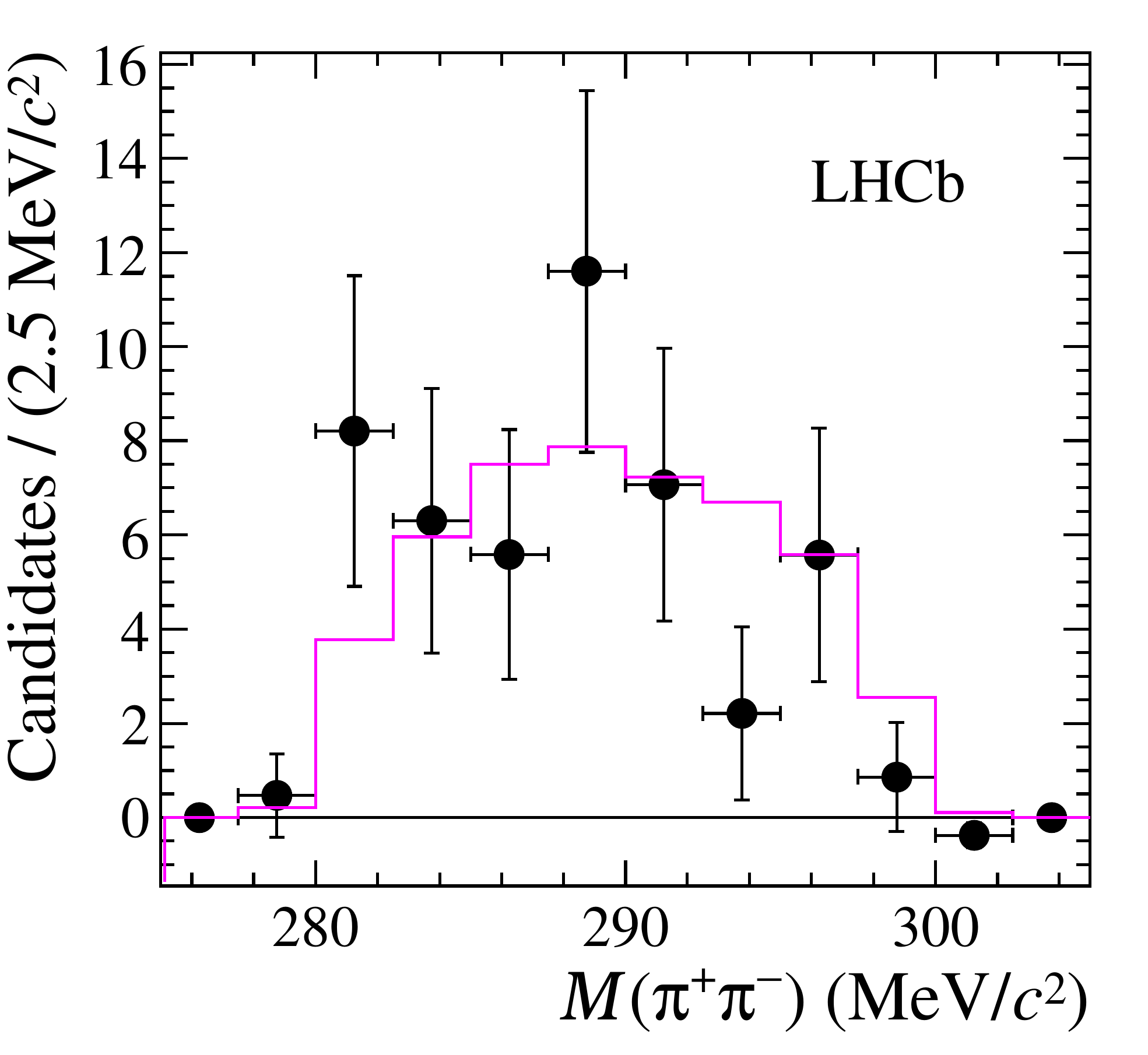}
  \end{center}
  \vspace*{-0.5cm}
  \caption{\small 
    Invariant mass of the two pions from $\lbh\to\lbpp$ decay. The points with the error bars are 
    background-subtracted data, the solid histogram is the result of phase-space decay simulation. }
  \label{fig:mhh}
\end{figure}

The invariant mass of the two pions, $M(\pi^+\pi^-)$, in the $\lbh\to\lbpp$ decay is shown in Fig.~\ref{fig:mhh}. 
The background is subtracted using the \sWeights procedure~\cite{splot}. The weights are calculated 
from the fit to \lbpp invariant mass distribution, which is practically uncorrelated with $M(\pi^+\pi^-)$. 
The $M(\pi^+\pi^-)$ distribution is consistent with the result of phase-space decay simulation, with 
$\chi^2/{\rm ndf}=1.6$ for ${\rm ndf}=9$. No peaking structures are evident. 

Systematic uncertainties on the mass measurement are shown in Table~\ref{tab:syst}. 
The dominant uncertainty in the absolute \lbst mass measurement comes from the uncertainty on the 
\Lb mass $\delta M_{\Lb}=0.69$~\mevcc; it is propagated to the \lbst mass uncertainty as 
$\delta M_{\lbst}=\delta M_{\Lb}\times(M_{\Lb}/M_{\lbst})\simeq 0.66$~\mevcc. 
This uncertainty mostly cancels in the mass difference $\Delta M_{\lbst}=M_{\lbst}-M_{\Lb}$, where
the residual uncertainty is $\delta\Delta M_{\lbst}=\delta M_{\Lb}\times(\Delta M_{\Lb}/M_{\lbst})$. 
The uncertainty of the signal parameterization is estimated 
by using the simulated signal parametrization without applying the resolution scaling factor, 
by using the natural width for both states when left free in the fit, 
and by conservatively including the \lblck contribution with the rate 12\% parameterized from simulation. 
The uncertainty due to the background parameterization is estimated by:
\begin{itemize}
\item using an alternative fit model for background description,
\item using the fit without the \lbppss constraint, 
\item using the fit with the background obtained from the simulation, 
\item fitting in the reduced invariant mass range 5910--5930~\mevcc, 
\end{itemize}
and taking the largest difference from the nominal fit result as systematic uncertainty. 
The effect of the momentum scale correction is 
evaluated by varying the scale coefficient by its relative uncertainty $5\times 10^{-4}$ in simulated signal samples. 

\begin{table}
  \caption{Systematic uncertainties on the mass difference $\Delta M_{\lbst}$ between \lbst and \Lb.}
  \label{tab:syst}
  \setlength{\extrarowheight}{2pt}
  \begin{center}
  \begin{tabular}{lcc}
    \hline
    \hline
    Source of        & \multicolumn{2}{c}{Systematic bias, MeV/$c^2$} \\
    uncertainty      & $\Delta M_{\lbl}$  & $\Delta M_{\lbh}$ \\
    \hline
      \Lb mass       & 0.034 & 0.035 \\
      Signal PDF     & 0.021 & 0.011 \\
      Background PDF & 0.002 & 0.002 \\
      Momentum scale & 0.008 & 0.013 \\
    \hline
      Total          & 0.041 & 0.039 \\
    \hline
    \hline
  \end{tabular}
  \end{center}
\end{table}

The significance of the observation of the two states is evaluated with simulated pseudo-experiments. A large number of 
background-only invariant mass distributions are simulated with parameters equal to the fit result, and 
each distribution is fitted with models that include background only, as well as background and signal. 
The mean mass value of the signal PDF is not constrained in the fit to account for a trial factor in the 
range 5900--5950~\mevcc.
The significance is calculated as the fraction of samples where the difference of the logarithms of fit likelihoods 
$\Delta\log\mathcal{L}$ with and without the signal is larger than in data. The fraction is obtained by 
an exponential extrapolation of the $\Delta\log\mathcal{L}$ distribution~\cite{gross} that allows a 
limited number of pseudo-experiments to be used for a signal with high significance. The significance 
is then expressed in terms of the number of standard deviations ($\sigma$). 
The significance of the \lbl state obtained in this way is $5.4\sigma$
for the $\Delta\log\mathcal{L}$ obtained from the nominal fit. To account for systematic effects, the minimum 
$\Delta\log\mathcal{L}$ among all systematic variations is taken; in that case the significance reduces
to $5.2\sigma$. Similarly, the statistical significance of the
\lbh state is 11.7$\sigma$, and the significance including systematic uncertainties is 10.2$\sigma$. 

The fit biases and the validity of the statistical uncertainties are checked with 
pseudo-experiments where the PDF contains both 
signal and background components. The fit does not introduce any noticeable bias on the measurement of the masses. The 
mass uncertainty for \lbh state is estimated correctly within 1\% precision; however, the mass uncertainty for the \lbl 
is underestimated by 4\%. This factor is taken into account in the final result. 

In summary, we report the observation of two narrow states in the \lbpp mass spectrum, \lbl and \lbh, with masses
\[
  \begin{split}
  M_{\lbl} & = 5911.97\pm 0.12 \pm 0.02 \pm 0.66 \mevcc, \\
  M_{\lbh} & = 5919.77\pm 0.08 \pm 0.02 \pm 0.66 \mevcc, \\
  \end{split}
\]
where the first uncertainty is statistical, the second is systematic, and the third is 
the uncertainty due to knowledge of the \Lb mass. 
The values of the mass differences with respect to the \Lb mass, where most of the last uncertainty 
cancels, and the remaining part is included in the systematic uncertainty, are
\[
  \begin{split}
  \Delta M_{\lbl} & = 292.60\pm 0.12(\mbox{stat})\pm 0.04(\mbox{syst}) \mevcc, \\
  \Delta M_{\lbh} & = 300.40\pm 0.08(\mbox{stat})\pm 0.04(\mbox{syst}) \mevcc. \\
  \end{split}
\]
The signal yield for the \lbl state is $17.6\pm 4.8$ events, and the significance 
of the signal (including systematic uncertainty and trial factor in the mass range 
5900--5950~\mevcc) is 5.2 standard deviations. For the \lbh state, the yield is $52.5\pm 8.1$ events and 
the significance is 10.2 standard deviations. The limits on the natural widths of these 
states are $\Gamma_{\lbl}<0.66$~\mev ($<0.83$~\mev) and $\Gamma_{\lbh}<0.63$~\mev ($<0.75$) at the 90\% (95\%) 
CL. 

The masses of \lbst states obtained in our analysis are 30--40~\mevcc higher than in the prediction 
using the constituent quark model~\cite{Garcilazo:2007eh}, and 20--30~\mevcc lower than the predictions based on 
the relativistic quark model~\cite{Ebert:2007nw}, modeling the color hyperfine interaction~\cite{Karliner:2008sv}
and an approach based on the heavy quark effective theory~\cite{Roberts:2007ni}. Calculation involving 
a combined heavy quark and large number of colors expansion~\cite{AzizaBaccouche:2001jc, AzizaBaccouche:2001pu}
gives a value roughly in agreement, although only the spin-averaged prediction is available. The earlier prediction 
based on the relativized quark potential model~\cite{Capstick:1986bm} matches well the absolute 
mass values for both states, but the \Lb mass prediction using this model is 35~\mevcc lower than the measured value. 

\vspace{\baselineskip}

We express our gratitude to our colleagues in the CERN accelerator
departments for the excellent performance of the LHC. We thank the
technical and administrative staff at CERN and at the LHCb institutes,
and acknowledge support from the National Agencies: CAPES, CNPq,
FAPERJ and FINEP (Brazil); CERN; NSFC (China); CNRS/IN2P3 (France);
BMBF, DFG, HGF and MPG (Germany); SFI (Ireland); INFN (Italy); FOM and
NWO (The Netherlands); SCSR (Poland); ANCS (Romania); MinES of Russia and
Rosatom (Russia); MICINN, XuntaGal and GENCAT (Spain); SNSF and SER
(Switzerland); NAS Ukraine (Ukraine); STFC (United Kingdom); NSF
(USA). We also acknowledge the support received from the ERC under FP7
and the Region Auvergne.

\addcontentsline{toc}{section}{References}
\bibliographystyle{LHCb}
\bibliography{main}

\end{document}